\documentclass{iaus}

\usepackage{graphicx}

\title[Interactive Solar Flare Modeling Tools]{New Interactive Solar Flare Modeling and Advanced Radio Diagnostics Tools} 

\author[Gregory D. Fleishman, Gelu M. Nita, \& Dale E. Gary ]   
{Gregory D. Fleishman$^{1,2}$, Gelu M. Nita$^{1}$
 \and Dale E. Gary$^1$}

\affiliation{$^1$New Jersey Institute of Technology, Newark, NJ 07102, USA
 \\[\affilskip]
$^2$
Ioffe Institute, St. Petersburg 194021, Russia \\ email: {\tt
gfleishm@njit.edu}
}

\pubyear{2010}
\volume{274}  
\pagerange{119--126}
 \jname{Advances in Plasma Astrophysics }
\editors{A. Bonanno, E. de Gouveia Dal Pino \&  A. Kosovichev, eds.}
\begin{document}

\maketitle

\begin{abstract}
The coming years will see routine use of solar data of unprecedented
spatial and spectral resolution, time cadence, and completeness in
the wavelength domain.  To capitalize on the soon to be available radio
facilities such as the expanded OVSA, SSRT and FASR, and the challenges
they present in the visualization and synthesis of the
multi-frequency datasets, we propose that realistic, sophisticated
3D active region and flare modeling is timely now and will
be a forefront of coronal studies over the coming years. 
Here we summarize our 3D modeling efforts, aimed at forward fitting of imaging
spectroscopy data, and describe currently available 3D modeling tools.  We also
discuss plans for future generalization of our modeling tools.

\end{abstract}

\firstsection 
\section{Introduction}

Solar activity, although energetically driven by subphotospheric
processes, depends critically on coronal magnetism, which,
broadly speaking, includes magnetic field generation, evolution, and
transformation into kinetic, thermal, and nonthermal energies in the
corona.  Reliable tools for doing direct diagnostics have been lacking,
although the situation is currently changing. Indeed, new space-
and ground-based solar optical telescopes are already capable of
precise measurements of the photospheric magnetic field with
sub-arcsecond angular resolution and high temporal resolution. When
combined with modern extrapolation algorithms, these data offer
important clues on the coronal magnetic field structure and
evolution. However, given the finite angular resolution,
sensitivity, observational errors, and even theoretical limitations, such
extrapolations are not unique, so the extrapolations require
independent verification.  An opportunity for quantitative
verification will be available when the new generation of
high-resolution solar-dedicated radio instruments (expanded OVSA,
SSRT, and FASR) become operational. Microwave radiation is
produced by the gyrosynchrotron (GS) mechanism as accelerated fast
electrons gyrate in the coronal magnetic field. As has been recently
proven using simulated microwave data, the coronal magnetic field can
indeed in principle be reliably recovered at the flare dynamic time
scales from the radio data, along with the key parameters of the
thermal plasma and accelerated electrons
(Fleishman et al. 2009). 
The ability to detect the magnetic field and its changes on dynamic
time scales is a critically needed element to uncover the
fundamental physics driving solar flares, eruptions, and activity.


\begin{figure} [htbp]
\includegraphics[height=1.8in]{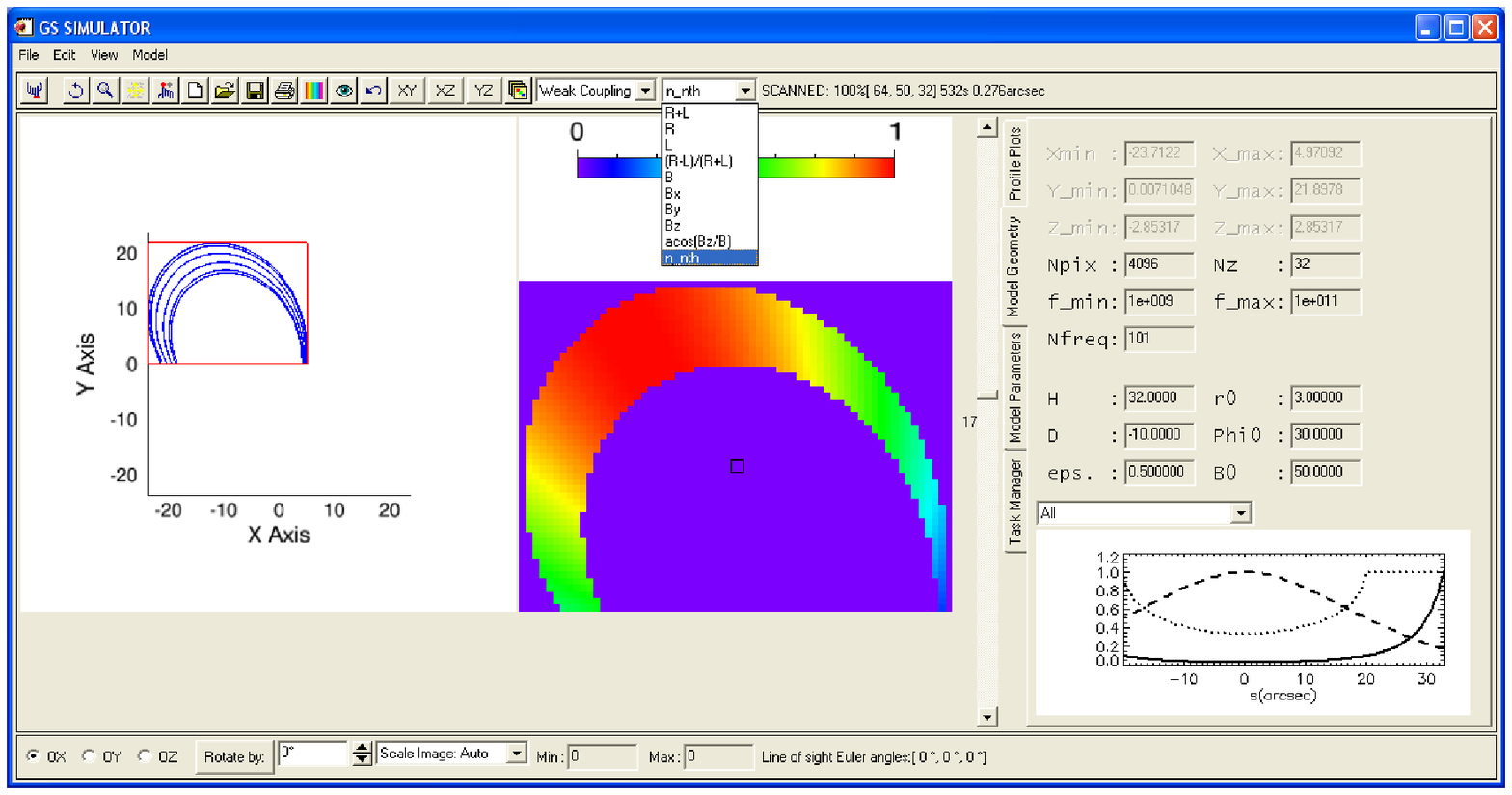}
\hspace{1cm}
\includegraphics[height=1.8in]{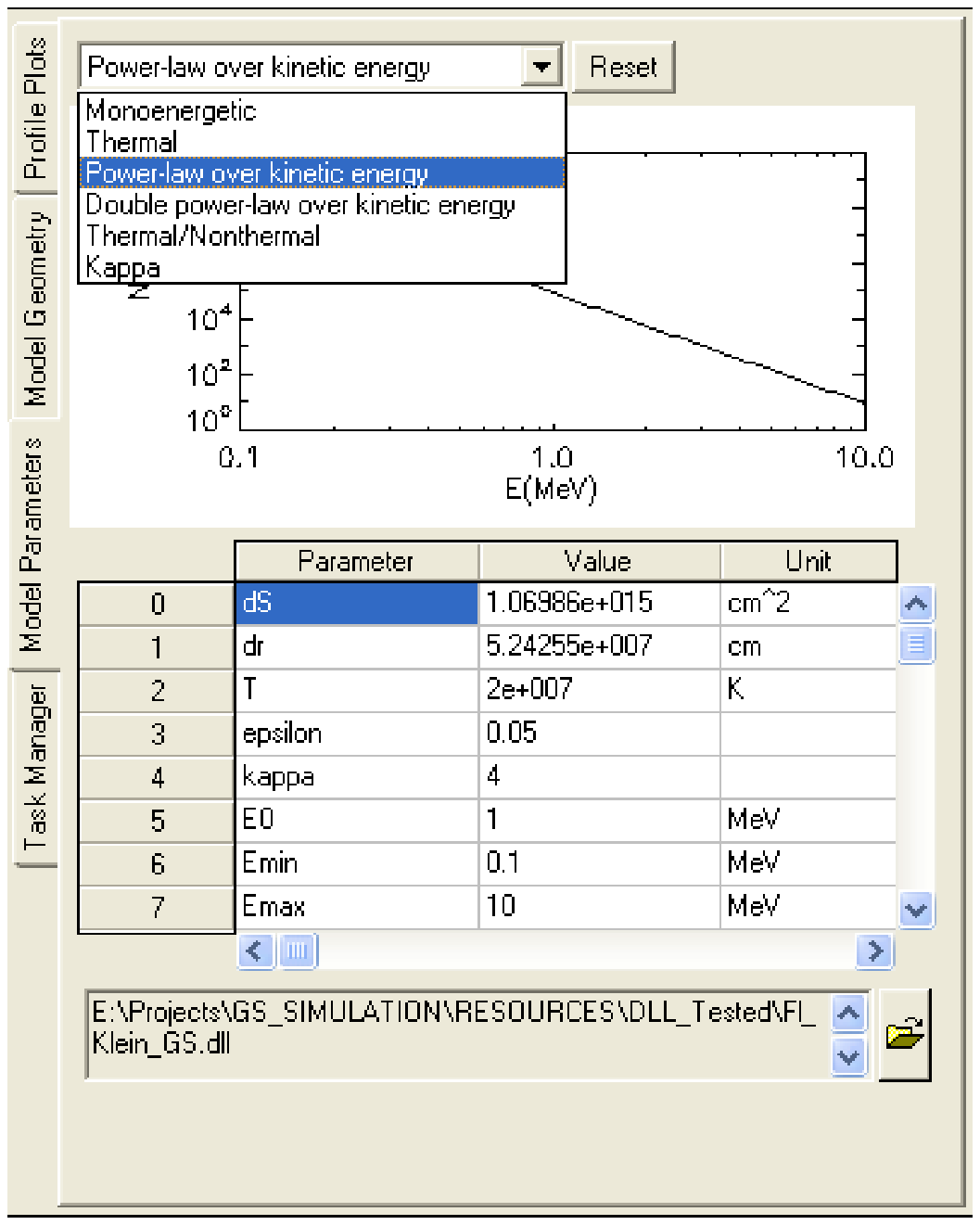}
\vspace{0.01in} \caption{\small \textsf{ Left: interface of the
model geometry and non-uniform parameter input---model magnetic loop
(left window); nonuniform distribution of fast electrons given in a
color code (middle window); and numeric input and graph display of
the parameters involved.  Right: selection of the fast electron
energy distribution from a number of pre-defined analytical
functions and parameters of this distribution.}}
\label{fig_GS_sim_2}
\end{figure}

\section{Modeling Methods and Tools}

\underline{\textbf{Direct Modeling.}} 3D models of the solar flares
are not yet numerous. Those available models 
(Preka-Papadema \& Alissandrakis 1992, Kucera et al. 1993, Bastian
et al. 1998, Sim{\~o}es \& Costa 2006, 2010, Tzatzakis et al. 2008,
Fleishman et al. 2009) are built on an idealized (e.g., dipole)
magnetic loop, rather than a realistic magnetic geometry.

Our currently available modeling tool, \textbf{GS Simulator}, is
also built based on an analytical (dipole) magnetic field model. A
flaring loop model (i.e., user specified dipole flux tube,
Figure~\ref{fig_GS_sim_2}, left) is produced with a newly developed
interactive IDL widget application intended to provide a flexible
tool that allows the user to generate spatially resolved GS spectra.
To do so, the user populates the loop with thermal plasma by
selecting plasma temperature and density in the adjustable parameter
list, Figure~\ref{fig_GS_sim_2}, right, and specifies a fast electron
population by selecting one of a few pre-defined distributions of
the electrons over energy and pitch-angle and choosing numeric
parameters for these distribution (Figure~\ref{fig_GS_sim_2}, right).
The spatial distribution of the electrons is also specified along the loop,
as shown in the color coded image (Figure~\ref{fig_GS_sim_2}, left).
After selecting the angle from which to view the loop with the mouse,
the tool then calculates the physical parameters along
each line of sight needed to solve the radiation transfer equation,
which is performed by external callable computing blocks. The
default codes generating the GS (and free-free) emission based on
the input geometrical line-of-sight model data were written in
FORTRAN and C++ based on fast GS codes newly developed by
\cite{Fl_Kuzn_2010},
and compiled as a DLL (or SO in case of Linux) callable by IDL.

\begin{figure}
\qquad\includegraphics[height=2.0in]{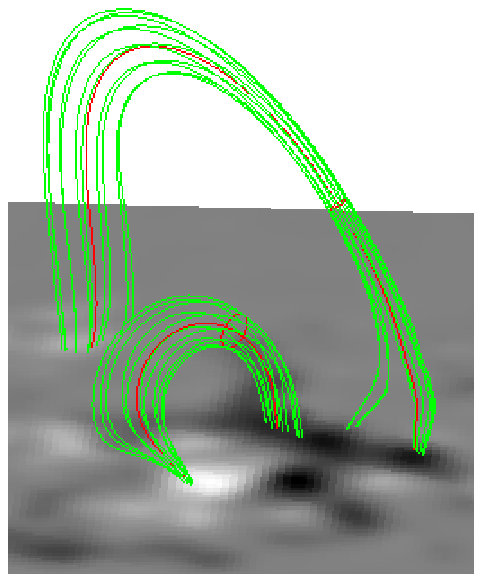} \hspace{0.51cm}
\includegraphics[height=2.0in]{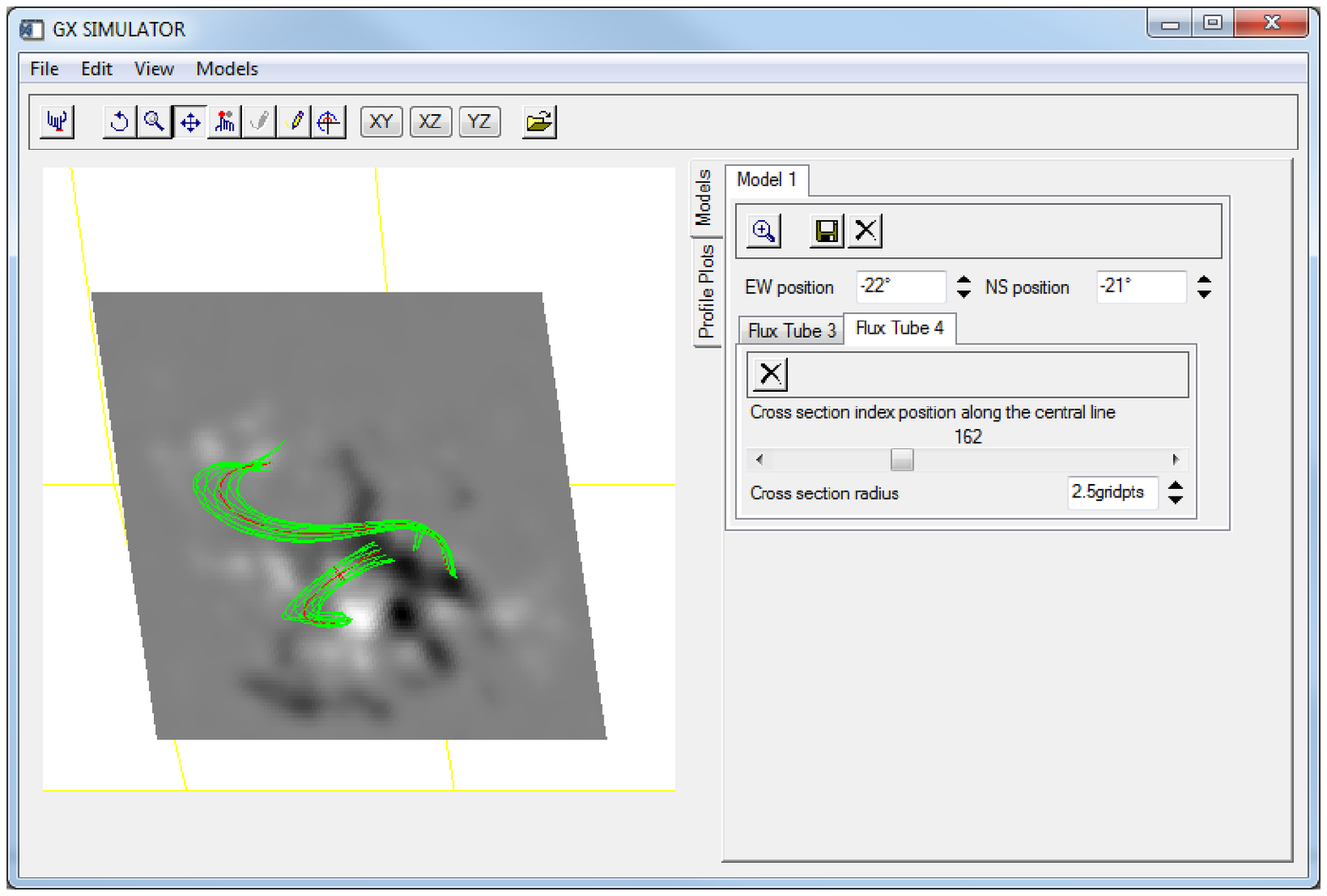}
\caption{\small\textsf{ Left: Two flux tube models. They can be
defined independently, and may represent interacting loops.
Additional controls are being prepared for populating the loops with
particles and ambient medium.  Even dynamics of the particles is
planned. Right: The loops now viewed from a specified location on
the solar disk (22 E 21 S in this example). The tabs labeled Flux
Tube 3 and Flux Tube 4 represent the two loops.}}\label{loops}
\end{figure}

The object-based architecture of this application provides the user
with full 3D interaction with a predefined, but adjustable, magnetic
loop geometry, as well as with any user defined analytical
geometrical model that would inherit the basic properties of the
generic "\verb"gs_model"" IDL object defined in this package. To
allow for more realistic 3D flare modeling, in place of drawing a
loop ``by hand", the tool must be further developed to include
(i) a numerical magnetic field structure, such as would be obtained
from a photospheric extrapolation (Figure~\ref{loops}) or a full MHD
model based on vector photospheric measurements of the magnetic and
velocity fields, along with the thermal plasma distribution, (ii) a
realistic electron acceleration and transport model, and (iii) the
ability to quickly compute emission in various wavelength regimes.
The tool will enable the user to identify, from the model or through
comparison with observations, the subset of field lines involved in
flaring, consider fast electron transport in this realistic magnetic
structure, calculate radiation from this 
evolving
volume, and so simulate a solar flare.



\underline{\textbf{Forward Fitting.}} The coronal
magnetic field is a key parameter controlling most solar flaring
activity, particle acceleration and transport. 
It has been understood, and often proposed, that the coronal
magnetic field along with fast electron distribution can in
principle be evaluated from the microwave GS radiation, which is
indeed sensitive to the instantaneous magnetic field strength and
orientation relative to the line of sight and to the fast electron
spectrum.

Diagnostics, understood as the determination of physical parameters
of a system under study from arrays of observed parameters, is a key
outstanding problem in Solar Physics. In some (basically linear)
cases 
regularized true inversions can work well (e.g., Kontar et al. 2004). 
In most of the cases, however, such true inversions fail because of
the highly nonlinear nature of physical systems. In such cases, the
forward fitting, i.e., finding a number of free parameters of a
physically motivated model of the system from fitting the model to
observations, can often be used successfully in place of true
inversions.

Anticipating a large breakthrough in the radio imaging spectroscopy
observations, which will become possible soon due to the next
generation of radio instruments, 
we have developed a practical forward fitting method, based on the
SIMPLEX algorithm with shaking, that allows reliable derivation of
the magnetic field and other parameters along a solar flaring loop
using microwave imaging spectroscopy of GS emission, which is
calculated with newly developed fast GS codes (Fleishman \&
Kuznetsov 2010). We illustrate the method using a model loop with
spatially varying magnetic field, filled with uniform ambient
density and an evenly distributed fast electron population with an
isotropic, power-law
energy distribution (Fleishman et al. 2009). 

\begin{figure}[!h]
\vspace{-0.153in}
\centerline{\includegraphics[height=1.1in]{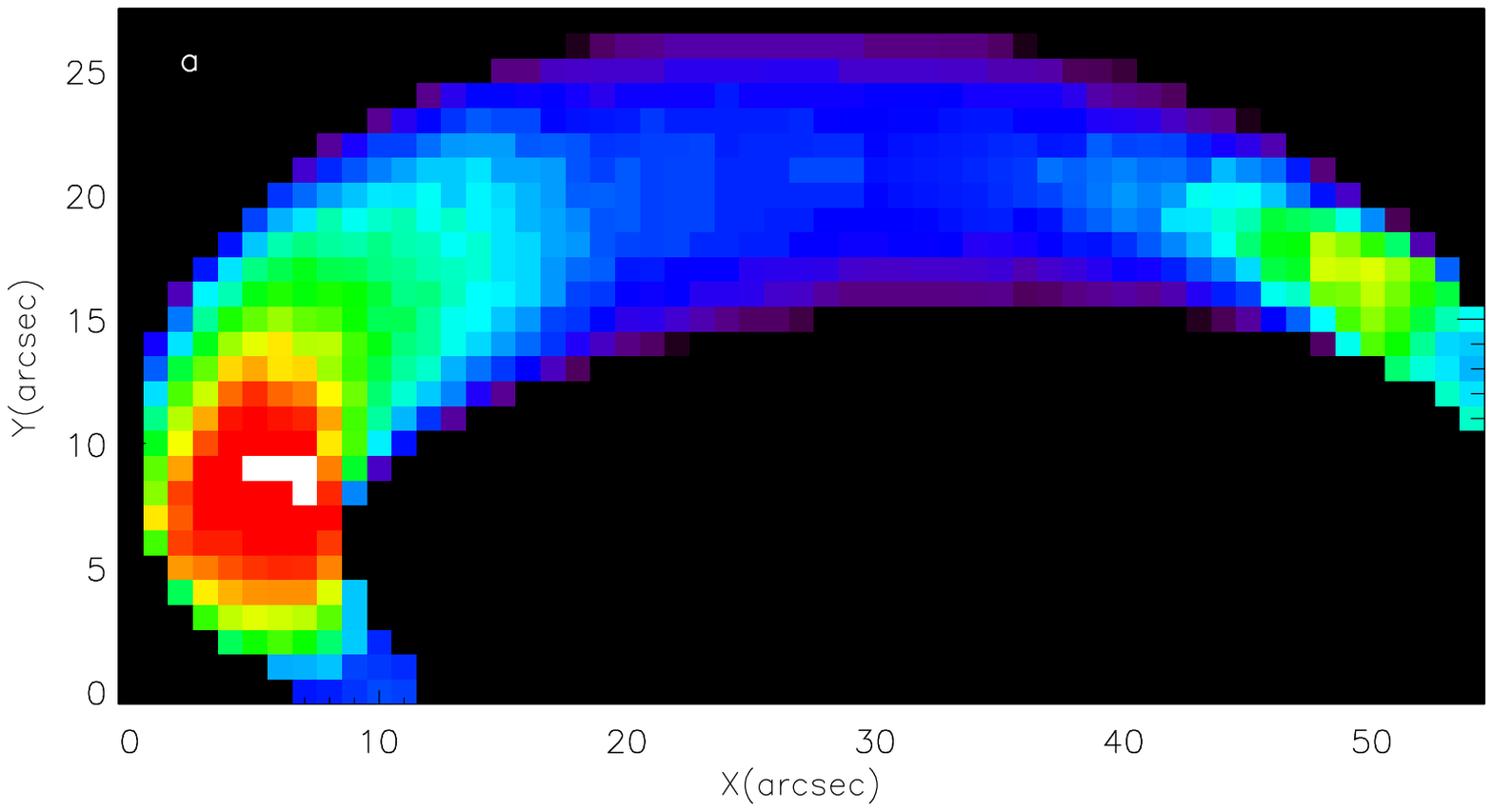}
\includegraphics[height=1.2in]{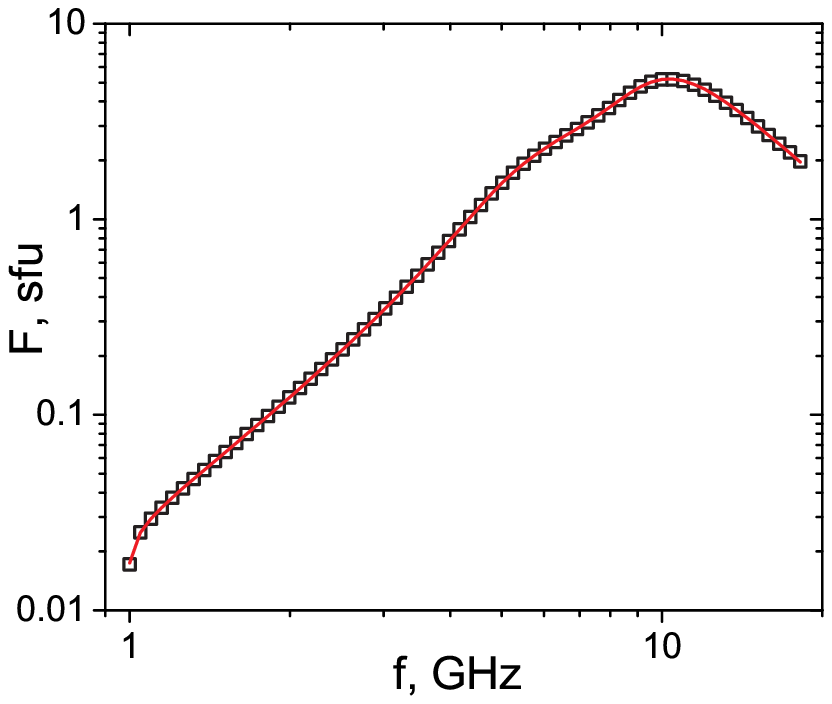}
\includegraphics[bb=24 527 278 702, height=1.1in,clip]{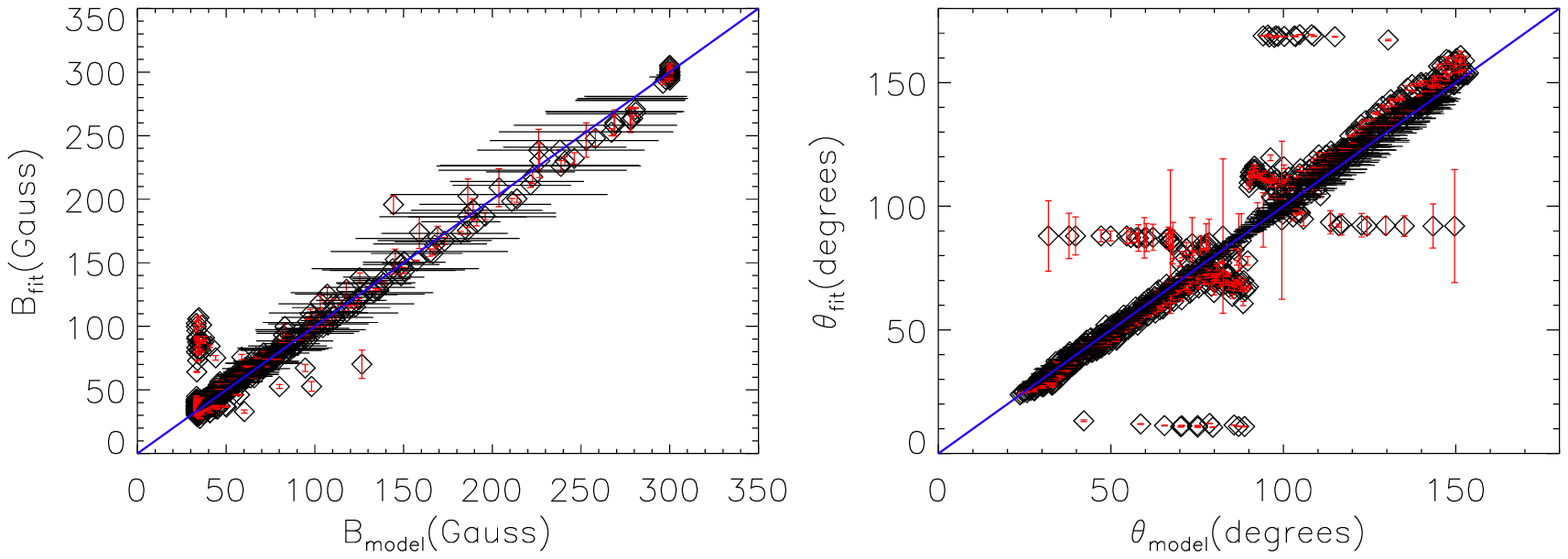}}
\caption{\label{image} {  \small \textsf{ Simulated image of the
radio emitting loop source at 4 GHz as observed by an ideal radio
heliograph with $\sim$1'' pixel size resolution (left).  Example of
the model (symbols) and fit (solid curve) spectra corresponding to
one particular pixel of the image displayed in the left panel
(middle). Model-to-fit comparison of the magnetic field (right).}}}
\end{figure}

From the flare radio model (Figure~\ref{image}, left) described
above we have a sequence of spatially resolved microwave spectra (in
a general case, both intensity and polarization data, one spectrum
per pixel). Then, we fit the data to a model microwave spectrum
pixel by pixel (Figure~\ref{image}, middle) to derive physical
parameters of the source (e.g., the magnetic field,
Figure~\ref{image}, right). Although the exact GS formulae are very
computationally expensive, much faster codes giving the same
accuracy have recently been developed by \cite{Fl_Kuzn_2010}, which
are used in practice as the forward fitting input. Then, having a
fitting procedure resulting in fast and reliable finding of the true
source parameters is exceedingly important. The problem here is that
most of the minimization algorithms often find a \emph{local}
minimum of the normalized residual (or of the reduced chi-square),
while the ultimate goal of the fitting  is to identify the
\emph{global} minimum. So far, we determined that the simplex
algorithm is very efficient in finding a local minimum. Then, it
needs to be 'shaken' for the simplex solution to overcome any local
minima and continue downhill towards the global minimum (a version
of the stimulated annealing approach). Even when the algorithm
performance is overall good, there is a non-zero probability that
the algorithm fails to find the true solution in some pixels. We use
post-processing to identify and flag/remove those pixels.

\section{Conclusions}


Modeling flare geometry, its full 3D visualization, and interactive
adjustments to the user-specific needs are highly complicated tasks
in themselves. Our modeling tools offer a united solution for these
problems, which are widely applicable to various external data cube
inputs and so offer a convenient framework for diverse studies of
coronal magnetism, including flares and the active region magnetosphere.
The modeling tools, computational libraries, and documentation are
available via author's web page, see
\verb"http://web.njit.edu/~gfleishm/".

The outlined modeling efforts can only bring fundamental knowledge
about flare/active region physics if used in conjunction with
modern, high-resolution observations. Key observations of the coronal
plasma parameters can only be made by radio instruments that combine
high sensitivity, temporal, spatial, and spectral resolution, which
are unavailable now. A small part of the required science will be
possible soon with the expanded OVSA instrument (anticipated operation
of the upgraded instrument begins in fall, 2013) and the upgraded
multi-wavelength SSRT. However, the full required capability has to
wait until the full FASR (Gary 2003) 
has been built.



\textbf{Acknowledgments.} This work was supported in part by NSF
grants ATM-0707319, AST-0908344, and AGS-0961867 and NASA grant
NNX10AF27G to New Jersey Institute of Technology, and by the RFBR
grants No. 08-02-92228, 09-02-00226, 09-02-00624. 


\bibliographystyle{apj} 
\bibliography{WP_bib,fleishman}

\end{document}